\documentstyle[sprocl,epsfig,psfig]{article}

\def\Journal#1#2#3#4{{#1} {\bf #2}, #3 (#4)}

\def\NPA{{\em Nucl. Phys.} A}
\def\NPB{{\em Nucl. Phys.} B}
\def\PLB{{\em Phys. Lett.}  B}

\def\PRD{{\em Phys. Rev.} D}

\def\beq{\begin{eqnarray}}
\def\eeq{\end{eqnarray}}

\begin{document}

\begin{flushright}
WU B 99-28 \\
hep-ph/9911253
\end{flushright}
\vskip 3.5\baselineskip

\title{$\gamma\gamma\to\pi^+\pi^-$ IN THE MODIFIED PERTURBATIVE APPROACH
\footnote{Talk given at the XVII Autumn School, ``QCD: Perturbative or Nonperturbative'', Lisbon, Portugal, 
29. Sept. - 4. Oct. 1999, to appear in the proceedings.}}

\author{C. VOGT}

\address{Fachbereich Physik, Universit\"at Wuppertal,\\ 42097 Wuppertal, Germany\\E-mail: cvogt@theorie.physik.uni-wuppertal.de}

\maketitle\abstracts{
We investigate the transverse momentum effects in $\gamma\gamma\to\pi^+\pi^-$ at moderately large total center of mass
energy $\sqrt{s}$ in the modified perturbative approach. The calculation of the differential cross section using
different approximations shows that the inclusion of $k_\perp$ in the quark and gluon propagators cannot lead
to a considerable improvement of the theoretical prediction.}

\section{Introduction}
For a long time it has been a matter of debate whether or not exclusive reactions are perturbatively dominated at
moderate momentum transfer. Calculations of various processes, the pion form factor being the most prominent, 
in the standard perturbative approach  \cite{BrodskyLepage} could not settle this question in a satisfactory manner.
The validity of this approach at scales of order of a few GeV has been questioned in Refs. \cite{IsgurSmith,Radyushkin},
where large endpoint contribution to form factors have been found. These large endpoint contributions render perturbative
calculations inconsistent, in the sense that observables receive their bulk contributions in domains where the strong coupling 
is rather large, thus destroying the applicability of perturbation theory. A more sophisticated investigation 
resulting in the modified perturbative approach including transverse momenta and Sudakov corrections lead to a self-consistent 
perturbative treatment of meson and nucleon elastic form factors 
\cite{BottsSterman,LiSterman,JakobKroll,BolzJakobKroll}. However, their results have clearly shown that the pion and 
proton form factors are not perturbatively dominated. Recent calculation of soft, overlap contribution to pion and proton 
form factors successfully describe the data at moderate momentum transfer \cite{JakobKrollRaulfs,Stefanis,DiehlFeldmann}.

Since the cross section of $\gamma\gamma\to\pi^+\pi^-$ has only been obtained in the standard perturbative approach, i.e. 
neglecting transverse momenta and Sudakov corrections, a new, improved investigation within the modified approach has been 
neccessary. The existing LO prediction \cite{BrodskyLepage} and NLO prediction \cite{Nizic} are far below the existing data.  
Moreover, these authors have used a rather large value for the pion form factor as a phenomenological input, 
which comes out much smaller in a self-consistent perturbative calculation.
The object of this work is to provide such an improved perturbative analysis of the process.

\section{The scattering amplitude}

Since the calculation of the transition amplitude for $\gamma\gamma\to\pi^+\pi^-$ in the modified approach is completely 
analogous to that of the pion form factor in Ref. \cite{JakobKroll}, we shall only very briefly introduce the objects
appearing in the hard scattering formula.\footnote{We implicitly assume that the running coupling and the
Sudakov factor are the same in the spacelike and timelike regions. We avoid further discussion of these delicate questions,
which are beyond the scope of the present work.} 

\begin{figure}[h]
\begin{center}
\psfig{file=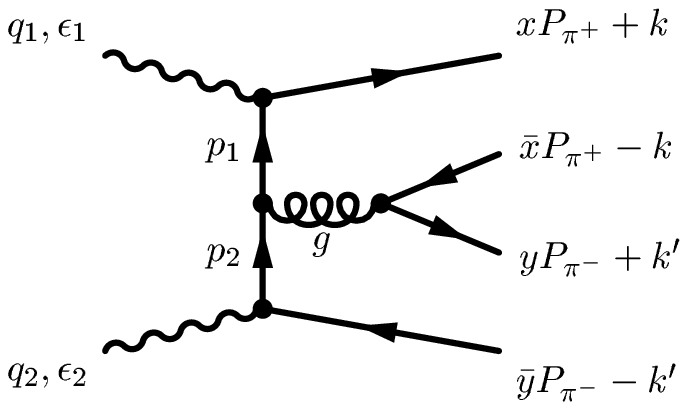,bb=180 650 420 790,width=5cm,angle=0}
\psfig{file=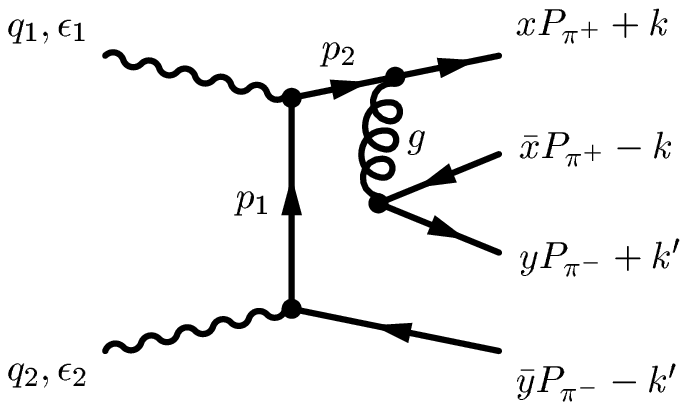,bb=180 650 420 790,width=5cm,angle=0}
\psfig{file=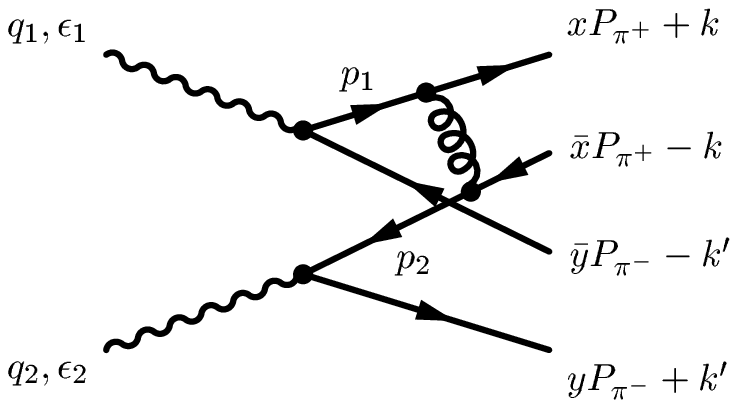,bb=180 630 420 790,width=5cm,angle=0}
\psfig{file=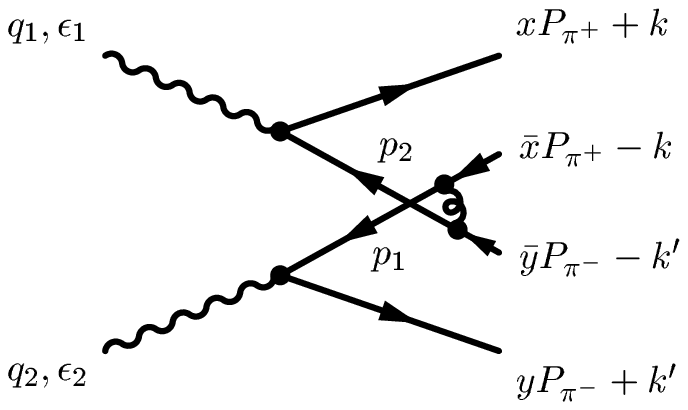,bb=180 630 420 790,width=5cm,angle=0}
\caption{The four basic diagrams for the $\gamma\gamma\to\pi\pi$ amplitude in the HSP. 
         We use the common notation $\bar{x}=1-x$.}    \label{graphs}
\end{center}
\end{figure}

The factorized amplitude for two gamma annihilation into pion pairs is given by a six dimensional phase space integral:
\beq
  {\cal M}^{\gamma\gamma\to\pi^+\pi^-}=&&\int dx dy \int \frac{d^2 b_\perp}{2 \pi^2}\frac{d^2 b'_\perp}{2 \pi^2}\;
   \hat{\Psi}_\pi(x,{\bf b}_\perp)\; \hat{T}_H(x,y,{\bf b}_\perp,{\bf b}'_\perp;s,t) \nonumber\\
    &&\times\;\hat{\Psi}_\pi(y,{\bf b}'_\perp)\; e^{-S},
\eeq
where $s$ and $t$ are the usual Mandelstam variables. $\hat{\Psi}_\pi$ denotes the Fourier transform of the 
light-cone wave function of the pion's valence Fock state and $\hat{T}_H$ is the Fourier transform of the 
hard scattering amplitude to be calculated perturbatively. The fractions $x,\, y$ describe how the quarks share 
their parents' pion momenta (cf. Fig. \ref{graphs}) and ${\bf b}_\perp$ and ${\bf b}'_\perp$ 
are the Fourier conjugated variables with respect to the intrinsic transverse momenta $k_\perp$ and $k_\perp'$. 
The factor $e^{-S}$ is the Sudakov form factor, which accounts for gluonic radiative corrections. 

Our ansatz for the pion wave function in transverse distance space reads
\beq
 \hat{\Psi}_\pi(x,{\bf b}_\perp)=\frac{f_\pi}{2\sqrt{6}}\;\phi_\pi(x,\mu_F)\,
      4\pi\exp\bigg[-\frac{x(1-x)}{4 a_\pi^2}b_\perp^2\bigg],
\eeq
where $f_\pi$ is the pion decay constant, $a_\pi$ is the transverse size parameter and $\mu_F$ is the factorization scale,
which is of the order of $\sqrt{s}$.
The functional form of the wave funciton meets various theoretical constraints and the parameters are fixed from normalization 
and certain pion decay processes respectivley. There is now broad agreement as to the distribution amplitude $\phi_\pi(x,\mu_F)$ 
being close to its asymptotic form $\phi_\pi^{AS}(x)=6\,x(1-x)$, which we shall employ in this work. 

The hard scattering amplitude $T_H$ at Born level is given by a total of 20 diagrams, which can be grouped together 
into four basic graphs displayed in Fig. \ref{graphs}. The remaining diagrams are obtained by various particle permutations.
Since we are only interested in the leading order behaviour in $k_\perp^2/s$ we neglect all transverse momenta
in the numerator of $T_H$. We need the Fourier transform of the hard amplitude which means we have to handle the 
product of two quark propagators and one gluon propagator. It is rather desirable to find an analytical expression 
of the Fourier transformed hard scattering amplitude in order not to lose too much numerical precision due to 
high dimensional numerical integration. This is a somewhat awkard task because each quark propagator
generally may depend quadratically and linearly on $k_\perp$ or $k'_\perp$ whereas the gluon propagators always depend on
both $k_\perp$ and $k'_\perp$. This means that the Fourier integral in general does not factorize with respect to $k_\perp$ 
and $k'_\perp$. Due to this circumstance one has to resort to certain approximations and at the same time make sure 
that important contributions from integration regions where quark and gluon propagators can go on-shell are not missed. 

At detailed numerical investigation has shown that the dominant effect of transverse momenta in the 
quark propagators is a considerable suppression at moderate values of $s$. This means that we can find an estimate from above 
of the cross section by using the collinear limit of the fermion propagators and keeping transverse 
momenta in the gluon propagators. The Fourier transformation then becomes trivial and we are generally left 
with a four instead of a six dimensional numerical integral. 

In order to make a more realistic approximation one should keep as many of the transverse momentum terms as possible,
such that an analytical evaluation of the Fourier transform of $T_H$ is still possible. In particular, one may neglect 
$k_\perp$ corrections of quark propagators in integration regions where the quark virtuality is spacelike, since in these 
regions the transverse corrections lead to a mild suppression of roughly $10\%$. 
A similar reasoning applies to gluon propagators. 

\section{Numerical results}

\begin{figure}[t]
\begin{center}
\psfig{file=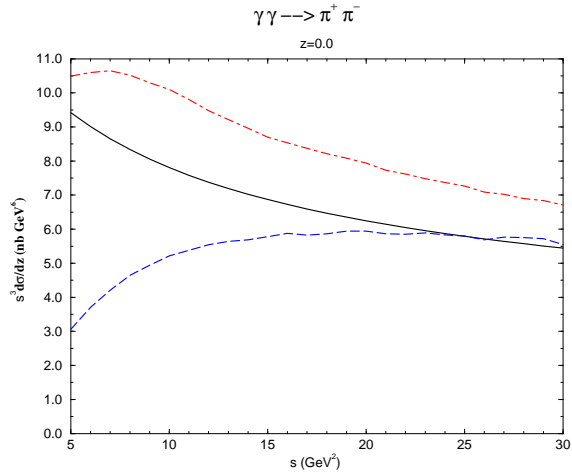,bb=45 70 530 675,width=6cm,angle=-90}
\caption{The differential cross section for $\gamma\gamma\to\pi^+\pi^-$ (scaled with $s^3$) at a center-of-mass scattering angle 
         $\Theta_{c.m.}=90^\circ$ using various approximations ($z=\cos\Theta_{c.m.}$). 
         The solid line shows the collinear case using a running coupling frozen at $\mu=1$ GeV. The dot-dashed line shows 
         the calculation where we have neglected transverse momenta in the quark propagators while  keeping them in the gluon 
         propagators. The dashed line represents the case where we have also taken into account transverse momenta in quark 
         propagators.} \label{dsdz}
\end{center}
\end{figure}
\begin{figure}[t]
\begin{center}
\psfig{file=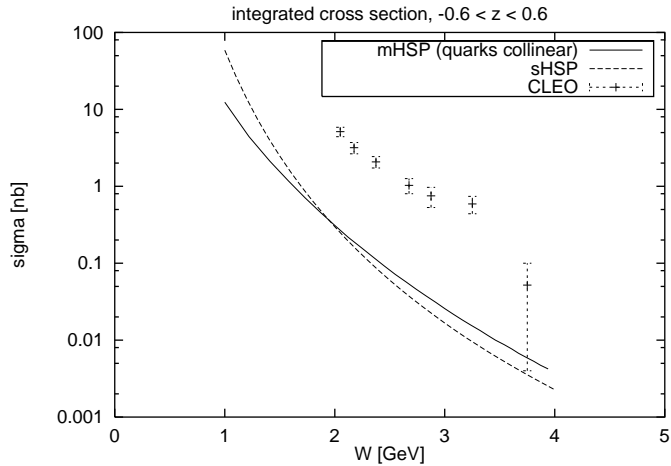,bb=40 50 550 410,width=9cm,angle=0} 
\caption{The integrated cross section for $\gamma\gamma\to\pi^+\pi^-$. The lines correspond to those of Fig. \ref{dsdz}. 
         Experimental data are taken from Ref. $^{12}$} \label{sigma}
\end{center}
\end{figure}

Results for the differential cross section are shown in Fig.\ref{dsdz}. We show both the approximations discussed above.
For comparison we also show the collinear case where we have used a running coupling
frozen at $\mu=1$ GeV. In Fig.\ref{sigma} we show our results for the integrated cross section. We have refrained
from the calculation of $\sigma_{\rm tot}(\gamma\gamma\to\pi^+\pi^-)$ in the case where we keep the transverse momenta 
in the fermion propagators, since, as we have discussed above, it cannot exceed the estimate using collinear fermion propagators.

We would like to emphasize that we have avoided to express the collinear amplitude in terms of the pion form factor.
Consequently, we have not explicitly assumed any phenomenological value of the pion form factor as has been done in Refs.
\cite{BrodskyLepage} and \cite{Nizic}. These authors have approximated the pion form factor by $F_\pi(s)=0.4$ GeV$^2/s$,
a value which is far too large from a perturbative point of view, since in the perturbative approach one obtaines 
a much smaller value. Such a large value as cited above can only be obtained by using a frozen coupling 
constant of the order of one over the whole integration region, a proceeding which is surely not consistent with the
application of perturbation theory. In fact, it is true that in order to explain the existing data of the 
$\gamma\gamma\to\pi^+\pi^-$ cross section one would need a rather large value for the pion form factor. A fit to these data 
even yields a value of $F_\pi(s)\approx 0.55$ GeV$^2/s$. However, such a large value is in no way justified by a perturbative 
calculation.

\section{Concluding remarks}

Our results show that even an improved perturbative analysis, i.e. taking into account transverse momentum corrections as 
well as Sudakov effects, does not 
lead to a considerable enhancement of the theoretical prediction of the $\gamma\gamma\to\pi^+\pi^-$ cross section.
Considering the fact that the collinear approximation of the cross section can be written in terms of the timelike
pion form factor, one might be tempted to expect an enhancement of even a factor four in the new calculation of the 
cross section, since the ratio of the timelike and spacelike form factor is of the order of two \cite{PireGousset} 
and the cross section depends quadratically on the form factor. However, it seems that in the collinear case the 
dependence of the cross section on the form factor is pure coincidence because if one includes transverse momenta 
in the calculation it is not possible to express the cross section in terms of the form factor, the reason being that
in the process under consideration the structure of quark and gluon propagators is generally quite different from the 
one appearing in the form factor.  

In view of this discussion, it is desirable to find a mechanism which accounts for nonperturbative contributions 
to two photon annihilation into meson pairs. We hope to provide further insights by future investigations in this
direction. 

\section*{Acknowledgements}
I would like to thank Th. Feldmann, R. Jakob and P. Kroll for helpful discussions as well as H. Anlauf for support with 
numerical problems. I am grateful to the DFG for financial support.

\end{document}